\documentclass[a4paper,twocolumn,superscriptaddress,showpacs]{revtex4}
\usepackage{graphicx}
\usepackage{dcolumn}
\usepackage{bm}

\usepackage[centertags]{amsmath}
\usepackage{amsfonts}
\usepackage{euscript}
\usepackage{amssymb}
\usepackage{amsthm}
\usepackage{newlfont}
\usepackage{stmaryrd}
\usepackage{mathrsfs}

\theoremstyle{plain}

\theoremstyle{definition}

\theoremstyle{remark}

\numberwithin{equation}{section}

  \let\de=\delta \let\ep=\epsilon
  \let\ga=\gamma 
   
  \let\th=\theta

\let\De=\Delta \let\Ga=\Gamma  \let\Om=\Omega

\newcommand{\caF}{{\mathcal F}}
\newcommand{\caG}{{\mathcal G}}

\newcommand{\caI}{{\mathcal I}}

\newcommand{\caW}{{\mathcal W}}

\newcommand{\opunit}{\text{1}\kern-0.22em\text{l}}

\newcommand{\bsP}{{\boldsymbol P}}

\DeclareMathAlphabet{\mathpzc}{OT1}{pzc}{m}{it}

\newcommand{\id}{\textrm{d}}

\begin{document}

\title{On the nonequilibrium relation between potential and stationary distribution\\ for driven diffusion}

\author{Christian Maes}
\email{christian.maes@fys.kuleuven.be}
\affiliation{Instituut voor Theoretische Fysica, K.U.Leuven, Belgium}
\author{Karel Neto\v{c}n\'{y}}
\email{netocny@fzu.cz}
\affiliation{Institute of Physics AS CR, Prague, Czech Republic}
\author{Bidzina M.~Shergelashvili}
\affiliation{Instituut voor Theoretische Fysica, K.U.Leuven, Belgium}

\begin{abstract}
We investigate the relation between an applied potential and the
corresponding stationary state occupation for nonequilibrium and
overdamped diffusion processes.  This relation typically becomes
long ranged resulting in global changes for the relative density
when the potential is locally perturbed, and inversely, we find
that the potential needs to be wholly rearranged for the purpose
of creating a locally changed density.  The direct question,
determining the density as a function of the potential,
comes under the response theory out of equilibrium. The inverse
problem of determining the potential that produces a given
stationary distribution naturally arises in the study of dynamical
fluctuations.
This link to the fluctuation theory results in a variational
characterization of the stationary density upon a given potential
and {\it vice versa}.
\end{abstract}


\maketitle
\section{Introduction}
Imagine independent colloidal particles in a potential field and subject to
friction and noise as imposed by a thermal reservoir or background
fluid. In thermal equilibrium at inverse temperature $\beta$,
Prob$[x] \propto \exp(-\beta V(x))$ where $V$ is the potential on
the states $x$. Typical examples include the Laplace barometric
formula but also the distribution of particles in a fluid
undergoing rigid rotation.  We now add an external forcing and we
wait till a steady regime gets installed. The stationary
statistics depends on the potential, but most certainly and
because of the forcing the resulting time-invariant distribution
of velocities and positions of the particles gets modified  with
respect to the Maxwell-Boltzmann statistics. The relation between
potential and stationary distribution is far from understood for
generic nonequilibrium systems, beyond its general specification
as being for example a solution to the time-independent
Fokker-Planck-Smoluchowski equation.

We show here that small local variations of the potential can globally affect the
relative density, provided a nongradient driving is present.  That effect
already arises in linear order around equilibrium.  For the inverse relation,
we are asking to reconstruct the potential which, under a known
driving, realizes a stationary distribution. In equilibrium, the
change in the potential needed to change the density
$\rho\longrightarrow \rho\,\exp(-A)$ is simply equal to that $A$.
Not so in nonequilibrium! It is then a genuine problem what
potential field can produce a given particle density at a fixed
driving.  Beyond obvious applications in interpreting
observational data, this question also naturally emerges in
dynamical fluctuation theory, cf.~\cite{MNW}. In particular, the
large fluctuations of certain time averages around their
stationary values are governed by a functional (sometimes called
an effective potential) that is given in terms of the potential of
the inverse problem. We will give two possible approaches to
finding the potential, one of which is analytical and the other
one is based on a variational formula.\\

The specific examples we treat in this paper and for which the
analysis can be applied are those of diffusing particles in a
background rotational velocity field. Diffusion in rotating media
is one of the central objects in geophysical and astrophysical
applications. The question of nonlocal and irreversible effects is
of particular interest for galactic dynamics, where according to
Chandrasekhar's theory the huge number of relatively small-size
gravitational encounters gives rise to an effective Brownian
motion for test stars, cf.~\cite{chan}; see also~\cite{bo} for an
account in the context of galaxy formation. Arguably, also
experimentally a most realistic scenario for maintaining a
constant nonconservative force is via differential rotation. One
can think of concentric cylinders rotating at different angular
frequencies which are imposed on the fluid by stick boundary
conditions. Far from equilibrium and under a general angular
(possibly angle-dependent) driving one can expect not only that
currents are being maintained but also that the time-symmetric
aspects of diffusion can be essentially changed. It is known that
the diffusion phenomenon itself may be influenced by rotation,
even by rigid rotation \cite{ppp}. Related considerations on
nonequilibrium diffusion also apply to other scenarios, including
e.g.~shear flow~\cite{foi,kata,re,san} and oscillatory
flows~\cite{knob}, and stochastic models of particle transport in
turbulent media necessarily include discussions of driven
diffusion phenomena, \cite{ga}. Colloidal particles in harmonic
wells and driven by shear flow have been explicitly treated for
the violation of the fluctuation-dissipation relation in
\cite{mau}.

The relation between
stationary density and potential is generically nonlocal or long
range but as our particles are mutually independent, that obviously says
nothing about the generic long range correlations in
nonequilibrium systems. As we will see however, the mathematical
mechanisms are not unrelated.  Various of these nonlocal effects
and generic long range correlations under nonequilibrium
conditions have been discussed in similar contexts of interacting
particle systems, most recently from the point of
fluctuation~theory~\cite{bod,DLS,jona}, in perturbation theory for
the stationary density~\cite{hal}, and as a result of the breaking
of the fluctuation-dissipation relation~\cite{gar,mkin,mr}.  It
was originally the mode-coupling theory in hydrodynamic studies
for a fluid not in thermal equilibrium that revealed the
(macroscopic) long-range correlations between fluctuations. Light
scattering experiments can reveal these correlations,
\cite{kirk,sen,seng,ls}.
In contrast, the present work analyzes the nonlocality on a
mesoscopic length scale and that does not result from particle
interactions but as is already present in a single particle
distribution (or, equivalently, in a density or an ensemble of
independent particles),
due to the imposed nonequilibrium driving.\\

In the next Section~\ref{2d} we specify our dynamical model: independent particles undergoing an overdamped diffusive motion in a confining potential and under driving. The nonlocality in the relation between potential and distribution is discussed throughout Section~\ref{non}.
The latter already speaks about the inverse problem of determining the potential for a given density, which is then elaborated on in Section~\ref{invp}, in the context of variational principles and for the purpose of dynamical fluctuation theory. In the Appendix we add further details about the McLennan's interpretation of nonequilibrium distributions, about the Green's functions encountered in the response problem, and about the dynamical fluctuation origin of the variational principles under consideration.

\section{Diffusion in a two-dimensional rotational fluid}\label{2d}

Restricting ourselves for the moment to the two-dimensional plane
with points labeled by the polar coordinates $x = (r,\th)$, we
consider an ensemble of independent test particles subject to a
rotation-symmetric force with potential $U(r)$, sufficiently
confining so that
\[
  Z = 2\pi \int_0^{+\infty} e^{-U} r\, \id r < +\infty
\]
The particles are suspended in a nonequilibrium fluid exerting an
additional force that can have some conservative component with
potential $\Phi(r,\th)$ and a nonconservative force $v = (v_r,
v_\th)$ for which we assume that the radial component $v_r = 0$
vanishes. The angular driving force $v_\th$ can be associated with
the local velocity of the background fluid which is maintained in
a differential rotation state. We assume a thermally homogeneous
background (setting the temperature to one), modeled by a Gaussian
and temporally white noise. Given that the motion is noninertial,
it satisfies the Langevin equation
\begin{equation}\label{mod}
  \id x_t = v\,\id t -\nabla (U + \Phi) \,\id t + \sqrt{2} \,\id B_t
\end{equation}
with $B_t$ standard two-dimensional Brownian motion.

An implicit assumption in the above construction is the smoothness
of the functions $U$, $\Phi$, and $v_\th$ on the domain taken here
as the entire two-dimensional plane. Yet, interesting  modifications arise
when the origin $r = 0$ is not accessible and the particles can only move
in the non-simply connected domain obtained by removing the latter.
This allows for possible singularities when approaching
the origin and the existence of a potential for the
driving force $v$ is no longer equivalent to the condition $\nabla
\times v = 0$; the rotational field of the form $v = (0,v_\th)$,
$v_\th \propto 1 / r$ serving as example. The exclusion of the
origin can be ensured e.g.\ by an infinitely repellent potential
therein or via suitable boundary conditions at the origin.

Note that in the absence of a potential, $U = \Phi = 0$, the
dynamics~\eqref{mod} does continue to make sense, although a
normalizable stationary distribution no longer exists. The
transient regime is still relevant and has been studied in
detail~\cite{rubi}.

The stationary distribution for the dynamics~\eqref{mod} has
density $\rho$ verifying the Fokker-Planck-Smoluchowski equation
\begin{equation}\label{cont}
  \nabla \cdot J = 0, \qquad
  J = \rho\,v - \rho \nabla(U + \Phi) - \nabla \rho
\end{equation}
We refer also to~\cite{duf1,san} for more thermodynamic and kinetic
gas considerations in the derivation of that nonequilibrium
dynamical equation. In polar coordinates the probability
current $J = (J_r, J_\th)$ takes the form
\begin{equation}\label{current-polar}
  J_r = -\rho\, \frac{\partial{(U + \Phi)}}{\partial r} -
  \frac{\partial\rho}{\partial r},\quad
  J_\th = \rho\, v_\th - \frac{\rho}{r} \frac{\partial\Phi}{\partial\th}
  - \frac{1}{r} \frac{\partial\rho}{\partial\th}
\end{equation}
By turning on the driving $v_\th$, typically not only an angular
current $J_\th$ is generated but also a nonzero radial component
$J_r$ does get maintained. This is \emph{a priori} not in
contradiction with the existence of a stationary distribution; its
normalizability essentially depends on the imposed potential $U +
\Phi$ and on the boundary conditions.

Our general aim is to analyze the relation between test potentials
$\Phi$ and stationary densities $\rho$, under a given confining
rotation-symmetric potential $U$ and as mediated by the rotational
field $v_\th$. First, we examine the issue of spatial nonlocality.

\section{Long-range response to changing the potential}
\label{non}

Nonlocal features have been widely discussed in the nonequilibrium
literature.  Mostly however that deals with the presence of long
range correlations, cf.~\cite{henk,kirk,duf,gar,jona,sasa} for
time-separated viewpoints, or with models of self-organized
criticality, cf.~\cite{gl}.  In our case, we have independent
particles, hence there are no correlations between the particles
and correlations between spatial points only appear because of
fixing the number of particles or by fixing the mass. We think of
the spatial dependence in the density as it is affected by local
changes in the external potential, and {\it vice versa}.\\

In the absence of driving, $v = 0$, the stationarity equation~\eqref{mod}
has the usual equilibrium solution
$\rho \propto e^{-U - \Phi}$, $J = 0$, which is manifestly a local functional of the test potentials $\Phi$ in the sense that the response
\[
  \frac{\de}{\de\Phi(z)} \Bigl[ \log  \frac{\rho(x)}{\rho(y)} \Bigr] =
  \de(z - y) - \de(z - x)
\]
is insensitive to perturbing $\Phi$ away from both points $x$ and $y$.

One can ask to what extent this is an equilibrium property but, in fact, it is easy to devise special nonequilibrium conditions where such a locality still holds. As a simple example, take $\Phi = 0$ and let the angular driving velocity be rotationally symmetric, $v_\th = u(r)$. Then, the stationary density and current become
\begin{equation}\label{ex1}
  \rho = \frac{1}{Z}\,e^{-U},\qquad
  J = (0, \frac{u}{Z}\, e^{-U})
\end{equation}
Although the stationary density coincides with the equilibrium solution and is perfectly local in the potential $U$, it now corresponds to a current-carrying steady state. Thus one cannot unambiguously decide just from the stationary distribution itself whether the system rests in equilibrium or whether irreversible flows are present.

Yet, generic nonequilibrium distributions do get modified due to driving and, as a result, they typically pick up some nonlocality. Next comes a simple demonstration.

\subsection{An exactly solvable model}\label{solvable model}

In the example above the radial currents are absent and the steady state is rotation-symmetric. A simple exactly solvable case where the rotation symmetry is broken can be obtained for an angular driving of the form
$v_\th = f(\th) / r$ and for test potentials that are constant along radials,
$\Phi(r,\th) = \Psi(\th)$. The origin is excluded by the boundary condition
$J_r(0) = 0$. In this case, the stationary density is found to be of the form
\begin{equation}\label{rr}
  \rho(r,\th) = p(r)\,q(\th),\qquad
  p(r) = \frac{1}{Z}\,e^{-U(r)}
\end{equation}
which is under the given assumptions the general form for a density with everywhere vanishing radial current,
$J_r = 0$. The steady state decomposes into separated concentric motions and the angular distribution $q(\th)$
in \eqref{rr} is determined from the stationarity condition~\eqref{cont}, which reads $J_\th(r,\th) = j(r)$ for some rotation-symmetric function $j(r)$ that can be determined. Explicitly, from~\eqref{current-polar},
\begin{equation}\label{separated}
  \{f(\th) - \Psi'(\th)\}\,q(\th) - q'(\th)
  = \frac{r j(r)}{p(r)}
\end{equation}
which decouples the polar coordinates and confirms the Ansatz \eqref{rr}. The angular current $j(r)$ is obtained by dividing~\eqref{separated} by $q$ and integrating over the angle variable, which yields, always
 for $v_\theta(r,\theta)= f(\theta)/r$,
\begin{equation}\label{ang-current}
  j(r) = \frac{p(r)}{r} \frac{\int_0^{2\pi} f(\theta)\,\id\th}{\int_0^{2\pi} q^{-1}(\theta)\,\id\th}
\end{equation}
As expected, a nonzero steady current is maintained whenever the angular driving does not allow for a potential, i.e., if the work performed over concentric circles is nonzero,
$w = \int_0^{2\pi} f\,\id\th \neq 0$. Using the normalization condition
$\int_0^{2\pi} q(\th)\,\id\th = 1$,
the solution of~\eqref{separated} is obtained in the form
\begin{align}\label{Gibbs-modified}
  q(\th) &= \frac{1}{\Om}
  \int_0^{2\pi} e^{ W(\th',\th)}\,\id\th',\,\,
\\
  \Om &= \int_0^{2\pi} \int_0^{2\pi}
  e^{W(\th',\th)}\, \id\th' \id\th \nonumber
\end{align}
with the work function
\[
  W(\th',\th) = \Psi(\th') - \Psi(\th)
  + \oint_{\th'}^{\th} f\,\id\xi
\]
Here we have used the notation $\oint_{\th'}^{\th}$ for the integral performed along the positively oriented path $\th' \rightarrow \th$ on the circle, i.e., it coincides with
$\int_{\th'}^{\th}$ for $\th' \leq \th$ whereas it equals to
$\int_{\th'}^{2\pi} + \int_0^{\th}$ otherwise. In the sequel we also employ the shorthand $\oint$ for the integral $\int_0^{2\pi}$.

Equivalently, the same solution to~\eqref{separated} can also be obtained in terms of the current $j(r)$; this leads to the next explicit expression for the latter:
\begin{equation}\label{current-explicit}
  j(r) = \frac{p(r)}{\Om\,r}
  \bigl( e^w - 1 \bigr)
\end{equation}

The equilibrium angular distribution is recovered for $w = 0$ (or
$j = 0$), in which case the work function $W(\th',\th)$ derives
from a potential and  formula~\eqref{Gibbs-modified} boils down to
the Boltzmann-Gibbs form.

For $w \neq 0$ the character  of the stationary density becomes
modified, as can be read from the response to changes in the test
potential $\Phi = \Psi(\theta)$ that we take to depend on the
angle $\theta$ only.  We take the functional derivative for
changes in the value of $\Psi$ at fixed angle $\eta$:
\begin{equation}\label{nonlocal-result}
  \frac{\de}{\de\Psi(\eta)} \Bigl[ \log
  \frac{\rho(r,\th)}{\rho(r,\th')} \Bigr] =
  Y(\eta,\th') - Y(\eta,\th)
\end{equation}
where
\[
  Y(\eta,\th) = \de(\eta - \th) -
  \frac{e^{W(\eta,\th)}}{\oint e^{W(\eta',\th)}\id\eta'}
\]
It is the second term that generates the nonlocality as its reciprocal
\begin{multline}
  \oint e^{W(\eta',\th)-W(\eta,\theta)}\id\eta'
\\
  = \int e^{\Psi(\eta) - \Psi(\eta')
  + \oint_\eta^{\eta'} \{ f - w\,\de(\xi - \th) \}\,\id\xi}\,\id\eta'
\end{multline}
still depends on $\th$.
Note also that since the
work function $W(\eta,\th)$ is discontinuous at $\th = \eta$, so
is the nonlocal contribution in the
response~\eqref{nonlocal-result} at $\eta = \th$ and $\eta =
\th'$.

That nonlocal term is manifestly a correction of order $O(w)$, hence, some more explicit information can be obtained within the
weak driving approximation, considering the driving force $f(\th)$ (or total work per cycle $w$) small. This yields, up to $O(w)$:
\begin{multline}\nonumber
  \frac{\de}{\de\Psi(\eta)} \Bigl[ \log
  \frac{\rho(r,\th)}{\rho(r,\th')} \Bigr] =
  \de(\eta - \th') - \de(\eta - \th)
\\
  + w\,q_*(\eta)
  \begin{cases}
    \oint_{\th'}^\th q_*\,\id\xi &
    \text{if } \eta \in (\th \rightarrow \th')
  \\
    \oint_{\th}^{\th'} -q_*\,\id\xi &
    \text{if } \eta \in (\th' \rightarrow \th)
  \end{cases}
\end{multline}
where $q_*(\eta)$ stands for the auxiliary density
\[
  q_*(\eta) = \frac{e^{\Psi(\eta)}}{\oint e^{\Psi} \id\xi}
\]
and $\eta \in (\th \rightarrow \th')$ indicates that $\eta$ belongs to the positively oriented path from $\th$ to $\th'$. This explicitly confirms the nonlocal and discontinuous structure of the response:
\emph{The relative density
$\rho(r,\th) / \rho(r,\th')$ of our weakly driven system
remains (strongly) sensitive to locally modifying the potential,
$\Psi(\th) \rightarrow \Psi(\th) + \ep\,\de(\th - \eta)$, at an arbitrary angle $\eta$.
The effect disappears at equilibrium, $w = 0$.}

Note that this special example is essentially one-dimensional and
we have so far only considered the response to an $r-$constant
change in the potential. The response to a strictly local
perturbation will be discussed in the next section by a more general
method.

The nonlocality of the inverse problem for this specific example, potential as function of the density, is continued around equations
\eqref{invl1}--\eqref{invl2}.

\subsection{General linear theory}\label{dire}

Having observed that the nonlocal features of nonequilibrium
states already emerge in the lowest order of the driving strength,
we can now follow the linear analysis more systematically. The
method is by now well known; we refer to earlier work of McLennan
and others, \cite{mac,Zub}.  The analysis starts from the
overdamped form~\eqref{mod}, and again for simplicity
we keep in mind diffusion
in the entire plane under natural boundary conditions (decay at infinity).

One takes the reference
equilibrium distribution (for $v=0$) as
\begin{equation}\label{39}
  \rho_o(x) = \frac 1{\cal Z}\,e^{-V(x)}
\end{equation}
where we combine $V=U + \Phi$ and $\cal Z$ is the normalization.
Assume now that the driving velocity field $v$ is uniformly small,
$|v| = O(\varepsilon)$.

The solution of the stationarity
equation~\eqref{cont} to linear order in $\varepsilon$ reads
\begin{equation}\label{ftc30}
  \log\frac{\rho}{\rho_o} =
  \frac{1}{L_V}\, (\nabla\cdot v - v \cdot \nabla V)
\end{equation}
where $L_V = -\nabla V \cdot \nabla + \De$, which is recognized as the
generator of a reversible diffusion in potential $V$; see Appendix~\ref{app: mclennan} for more details and for a physical interpretation in terms of the McLennan's theory, \cite{mac}. Writing equation~\eqref{ftc30} in the form
\begin{equation}\label{sgn}
  \Delta \nu - \nabla V\cdot\nabla \nu =
  \nabla\cdot v - v\cdot \nabla V
\end{equation}
with $\nu = \log (\rho/\rho_o)$, its variation along $V
\longrightarrow V + \delta V$ is
\[
\Delta \delta \nu   - \nabla V\cdot \nabla \delta\nu - \nabla
\delta V \cdot \nabla\nu = -v\cdot\nabla\delta V
\]
In terms of the equilibrium generator $L_V$ and the stationary
current $J = \rho(v - \nabla V) - \nabla \rho$, this reads
\begin{equation}\label{pois}
  L_V\, \de\nu = -\frac{J}{\rho} \cdot \nabla \de V
\end{equation}
always up to terms $O(\varepsilon^2)$; in the same order the density $\rho$ on the right-hand side can be replaced by $\rho_o$.
This is a Poisson equation for $\delta\nu$ in which the Laplacian (free
diffusion) is modified with a drift term in potential $V$.
Note that since
$L_V = \frac{1}{\rho_0}\,\nabla \cdot (\rho_0 \nabla)$ and
$\nabla \cdot J = 0$, this problem is equivalent to studying the electrostatic potential in an inhomogeneous dielectric environment as generated by a source with zero total charge. Whereas the local component of the response to $\de V$ is already hidden in the $V-$dependence of the equilibrium density $\rho_o$, the nonlocal character of solutions to~\eqref{pois} follows from general features of elliptic operators.
The solution of the free Poisson equation (with only the Laplacian and natural boundary conditions at infinity) is explicit and manifestly long range. The potential $V$ or the finiteness of the system introduces an extra confinement and the claim needs to be refined. One expects that within the confinement region where the density $\rho_o$ is approximately constant and near its maximal value, the response $\de\nu(x)$ at $x$ to a local perturbation $\de V$ (both localized within that region) can be well estimated by replacing $L_V$ with the free Laplacian $L_0 = \De$. This intuition is indeed correct as we shortly explain in Appendix~\ref{ap}.
In that region, \emph{the response $\de\nu(x)$ to a local perturbation
$\de V$ derives from the Green's function for $d-$dimensional Brownian motion, weighted by the local mean velocity $J/\rho$.} In this sense the nonlocal response is intrinsically a nonequilibrium feature.\\

Although the above linear analysis can formally be extended far from equilibrium and one obtains a generalization of the response formula~\eqref{pois}, the linear operator replacing $L_V$ is no longer symmetric, in agreement with the breakdown of Onsager reciprocity relations. We will see in the next section that an inverse formulation of our linear response question does not suffer from the above complication and it also becomes remarkably easier to study outside the weak driving regime.

\subsection{Local response to nonlocal perturbation}
\label{invnon}

In this section we discuss long range aspects
in the inverse problem, namely, how the test potential $\Phi$ that makes a \emph{given} density $\rho$ stationary, is affected by a local change in $\rho$. The relevance of this question and some other approaches are considered in~Section~\ref{invp}.\\

We start by revisiting the exactly solvable model of Section~\ref{solvable model}. The confining potential
$U(r)$ and the driving $v(r,\th) = (0, f(\th) / r)$ are always considered fixed. For the class of radial-angular uncorrelated densities,
$\rho(r,\th) = p(r)\, q(\th)$, the stationary equation~\eqref{cont} is satisfied for the test potential of the form
$\Phi(r,\th) = \Phi_0(r) + \Psi(\th)$, with the radial part
\begin{align}\label{invl1}
  \Phi_0 &= -\log p - U
\\ \intertext{and with the angular part equal to}
  \Psi &= -\log q + \int_{\th_0}^\th
  \Bigl( f - \frac{C}{q} \Bigr)\,\id \xi
\end{align}
Here $\th_0$ is an arbitrary fixed angle and the constant $C$ is determined from
\begin{equation}\label{invl2}
  C = \frac{\oint f\,\id\th}{\oint q^{-1} \id\th}
\end{equation}
The latter specifies the corresponding stationary current field as
$J = (0, C / r)$. Away from equilibrium, i.e., for $w = \oint f\,\id\th \neq 0$, the angular component $\Psi(\th)$ apparently becomes a nonlocal functional of the angular density
$q(\th)$, similarly to what we have observed for the functional dependence $\rho[\Psi]$.\\

In the general case, we are asked to find the potential $\Phi$
that solves the stationarity equation \eqref{cont} as a function
of density $\rho$ for given velocity field $v$ and confining
potential $U$, say on $d-$dimensional space. By changing $\rho \longrightarrow
\rho + \de\rho$ with $\int \de\rho(x)\,\id x = 0$, we get the
linear response equation for $G = U + \Phi + \log\rho$,
\begin{equation}\label{diff}
  \frac 1{\rho}  \nabla \cdot (\rho \nabla\de G) =
  \frac{J}{\rho} \cdot \nabla \de\nu
\end{equation}
with $\nu = \log\rho$ and $J = \rho(v - \nabla G)$ the stationary current.  We recognize in the left-hand side of \eqref{diff} the action of the
generator
\[
  L_\rho f(x) = \Delta f(x) + \nabla \log \rho \cdot \nabla f
\]
for the function $f=\delta G$.  The linear operator $L_\rho$
generates a \emph{reversible} diffusion in potential $-\log \rho$, i.e.\ an equilibrium process.  The analysis is now exactly similar as from
\eqref{pois}, but with $\nu$ replacing $-V$, and now restricting to a confinement region where $\rho(x)$ is approximately constant and maximal.
The source is nonzero by the
presence of $J\neq 0$ in the right-hand side of \eqref{diff}. We
refer again to Appendix~\ref{ap} for the analysis, but the
conclusion remains that
\emph{a generic local change in density $\de\rho$ in the confinement region requires a nonlocal adjustment of the potential and the corresponding response function derives from the Green's function for the $d-$dimensional Laplacian.}

Remark that no weak driving or small current assumption was employed in the above argument! In fact, the presence of the (auxiliary) reversible diffusion generated by $L_\rho$ suggests that, even far from equilibrium, there are symmetries in the response functions. This is indeed true, see Appendix~\ref{app: reciprocity} for such a reciprocity relation.

\section{More about the inverse problem}\label{invp}

One can ask how to actually construct the test potential $\Phi$
that makes a \emph{given} density $\rho$ stationary, i.e., solving~\eqref{cont}. An immediate application is found in dynamical
fluctuation theory; a brief review is left to
Appendix~\ref{dyflu}. In the following we present a general
procedure to solve the inverse stationary problem for a class of
densities. Next, a variational formulation suitable for numerical
implementation will be given.

\subsection{A general solution}
We are back to the set-up of Section \ref{2d} for two-dimensional
rotational diffusion. We restrict to those densities $\rho$ that
are everywhere bounded from zero and for which all (nonempty)
equilevel lines, $\rho(r,\th) = \text{const}$, are closed curves. We also
stick to trivial boundary conditions at infinity, as guaranteed by
a sufficient decay of all the fields $\rho$, $U$, and $v$. The
auxiliary velocity
\begin{equation}\label{c-introduced}
  c = v - \nabla(U + \Phi + \nu),\qquad
  \nu = \log\rho
\end{equation}
shares with  the driving field $v(r,\th)$ an equal vorticity,
\begin{equation}\label{c-rot}
  \nabla \times c = \nabla \times v
\end{equation}
The probability current is $J = \rho\,c$, and the stationarity condition~\eqref{cont} reads
\begin{equation}\label{c-div}
  \nabla \cdot c + c\cdot \nabla\nu = 0
\end{equation}
We only need to find the vector field  $c(r,\th)$; then the
potential $\Phi(r,\th)$ can be calculated as
\begin{equation}\label{Phi-explicit}
  \Phi(x) = -\nu(x) - U(x) +
  \int_{\ga:\, x_0 \rightsquigarrow x} (v - c) \cdot \id \ell
\end{equation}
modulo a constant, where the integral is taken along an arbitrary
curve connecting a fixed initial point $x_0$ with $x$.

To determine $c(r,\th)$ solving equations~\eqref{c-rot}--\eqref{c-div}, we first observe that it is unique by the Helmholtz decomposition theorem when supplying the boundary condition that the difference $c - v$ goes to zero at infinity. Still another boundary condition has to be added in the case the origin is not accessible and excluded from the domain.

In the following we restrict ourselves again to
the two-dimensional plane.
Equation~\eqref{c-div} is solved by any vector field of
the form
\begin{equation}\label{c-general}
  c(r,\theta) = g(\nu(r,\theta))\, \Bigl(-\frac{1}{r}
  \frac{\partial \nu}{\partial \theta},
  \frac{\partial \nu}{\partial r} \Bigr)
\end{equation}
with $g$ an arbitrary function. The latter is fixed by condition~\eqref{c-rot},
\begin{equation}\label{eqrotat}
  \nabla \times \Bigl\{ g(\nu(r,\theta))\, \Bigl(
  -\frac{1}{r}\frac{\partial \nu}{\partial \theta},
  \frac{\partial \nu}{\partial r}\Bigr) \Bigr\} =
  \nabla \times v
\end{equation}
that after integration over the surface enclosed by any equilateral curve of the density, $\nu(r,\th) = a$, and using Stokes' theorem yields
\begin{equation}\label{sto}
  \oint_{\nu = a} v \cdot \id\ell =
  g(a) \oint_{\nu = a}
  \Bigl( -\frac{1}{r}
  \frac{\partial \nu}{\partial \theta},
  \frac{\partial \nu}{\partial r} \Bigr) \cdot \id\ell
\end{equation}
Parameterizing the curve $\nu(r,\th) = a$ by its proper length so that
\[
  \id \ell = \Bigl[ \frac{1}{r^2}
  \Bigl(\frac{\partial \nu}{\partial \theta} \Bigr)^2 +
  \Bigl(\frac{\partial \nu}{\partial r} \Bigr)^2 \Bigr]^{-1/2}
  \Bigl( -\frac{1}{r} \frac{\partial \nu}{\partial \theta},
  \frac{\partial \nu}{\partial r} \Bigr) \,\id s
\]
we finally get
\begin{equation}\label{g-explicit}
  g(a) = \frac{\oint_{\nu = a} v \cdot \id\ell}{\oint_{\nu = a}
  \bigl[ \frac{1}{r^2}
  \bigl(\frac{\partial \nu}{\partial \theta} \bigr)^2 +
  \bigl(\frac{\partial \nu}{\partial r} \bigr)^2 \bigr]^{1/2}\,\id s}
\end{equation}
Formulae~\eqref{Phi-explicit}, \eqref{c-general}, and \eqref{g-explicit} together provide an explicit solution for the test potential $\Phi$.

As a check we take the example \eqref{ex1}; there
\[
  c = (0, u(r)) = v
\]
and the curves $\nu = a$ are concentric, corresponding to the
equipotential lines for $U(r)$, assuming that it is monotone in
$r$.  Therefore \eqref{g-explicit} gives $g(a) = -u(r) / U'(r)$ for
$a = -U(r) - \log Z$.

Further, the more general example of Section~\ref{solvable model} has
$c = (0,j(r)/q(\theta))$ with  non-vanishing divergence whenever $q$ is
not a constant, in contrast with the form \eqref{c-general}. The
point is that this example has a velocity
field which is  not defined at the origin, it being excluded from
the domain. Hence, Stokes' theorem in the form~\eqref{sto} cannot
be used and a modification is needed; we omit details.\\

The above solution, basically obtained by a suitable deformation
of polar coordinates,
provides a class of examples with non-vanishing radial current.
The latter is generally the case whenever $\nu$ does not decompose into independent radial and angular parts; compare with the model of Section~\ref{solvable model}.

\subsection{Variational approach}
\label{variational}

Write now the Fokker-Planck-Smoluchowski equation~\eqref{cont},
considered again as the inverse stationary problem for the test
potential $\Phi$, in the form
\begin{equation}\label{poisson}
  \nabla \cdot (J_0 - \rho \nabla \Phi) = 0
\end{equation}
with $J_0 = J_0(\rho)$ the $\Phi-$independent part of the probability current $J$:
\begin{equation}
  J_0(\rho) = \rho(v -\nabla U) - \nabla \rho
\end{equation}
Recalling that it is an elliptic partial differential equation for
$\Phi$, its solution coincides with the minimizer of the quadratic
functional
\begin{equation}\label{td}
  \cal F_\rho[\Psi] = \frac{1}{2}\int \rho \nabla\Psi \cdot
  \nabla\Psi\,\id x
  + \int \Psi \,\nabla \cdot J_0(\rho)\,\id x
\end{equation}
under the unchanged boundary conditions if present. This formulation is suitable for numerical computations, see e.g.\ via \cite{finel}.

There are other variational principles of physical importance that appear intimately related to our inverse stationary problem. To explain those, consider the functional
\begin{equation}\label{G-explicit}
  \caG[\mu,j] = \frac{1}{4} \int \mu^{-1} [j - J_0(\mu)] \cdot
  [j - J_0(\mu)]\,\id x
\end{equation}
defined for all normalized densities $\mu$ and all divergenceless currents
$j$, $\nabla \cdot j = 0$; see Appendix~\ref{dyflu} for its meaning within the dynamical fluctuations theory. This functional is manifestly positive and zero only if $\mu$ and $j$ coincide with the stationary density respectively the stationary current (for the case $\Phi = 0$). It can be used to construct the variational functional
\begin{equation}\label{I-variational}
  \caI[\mu] = \inf_{j:\,\nabla \cdot j = 0} \caG[\mu,j]
\end{equation}
with minimizer equal to the stationary density.
This is a constrained variational problem that can be solved by Lagrange multipliers; the solution reads
\begin{equation}\label{I-explicit}
  \caI[\mu] = \frac{1}{4} \int \mu \nabla\Phi \cdot \nabla\Phi\,\id x
\end{equation}
with $\Phi$ the test potential that makes the density $\mu$ stationary, cf.~\eqref{poisson},
\begin{equation}\label{poisson-again}
  \nabla \cdot [J_0(\mu) - \mu\nabla\Phi] = 0
\end{equation}
In this way, the solution to the inverse stationary problem is an essential step in constructing the variational functional $\caI[\mu]$ on densities.

Finally, recall that equation~\eqref{poisson-again} is equivalent to the variational problem $\caF_\mu[\Psi] = \min$ with $\caF_\mu$ introduced in~\eqref{td}. Combining with~\eqref{I-explicit} we have the relation
\begin{equation}
  \inf_\Psi \caF_\mu[\Psi] = \caF_\mu[\Phi]
  = -2\caI[\mu]
\end{equation}
that yields the next expression for the functional $\caI[\mu]$ (changing $\Psi \longrightarrow 2\Psi$ for convenience and integrating by parts):
\begin{equation}\label{I-DV}
  \caI[\mu] = \sup_\Psi \int \nabla\Psi \cdot [J_0(\mu) - \mu\nabla\Psi]\,
  \id x
\end{equation}
This \emph{unconstrained} variational formula is apparently more useful for numerical computations than~\eqref{I-variational} above.

\section{Conclusion}

The relation between potential and stationary density in
mesoscopic (stochastic) systems appears to be generically long
ranged whenever there is a true nonequilibrium driving.
That long range effect is \emph{a priori} distinct from the long
range correlations under conservative dynamics extensively studied
before, and it occurs already for free particles. In models of
overdamped diffusions considered in this paper, we have linked
that long range effect to the slow spatial decay of the Green's
function for a certain equilibrium diffusion process (in the first
order around equilibrium).

{\it Vice versa}, a similar nonlocal change of potential is
generically needed to create a local change in the stationary
density. This issue appears relevant for the inverse stationary
problem that naturally emerges in the context of dynamical
fluctuations and nonequilibrium variational principles. We have
indicated how these specific issues become mutually related,
together with comparing some numerically feasible schemes based on
the dynamical fluctuation theory that might be of use in
applications.

\begin{acknowledgments}
K.~N.\ acknowledges the support from the project
AV0Z10100520 in the Academy of Sciences of the Czech Republic and from the Grant Agency of the Czech Republic (Grant no.~202/07/0404). C.~M.\ benefits from K.U.Leuven grant OT/07/034A.
\end{acknowledgments}

\appendix

\section{McLennan theory of stationary distributions}\label{app: mclennan}

It is useful to see how the original McLennan reasoning~\cite{mac} can be used to provide some physical interpretation for the stationary density of a weakly driven diffusion. We can write formula~\eqref{ftc30} in the equivalent form, always to linear order in $\ep$,
\begin{equation}\label{ftc3}
  \rho(x) = \rho_o(x)\,\exp\int_0^{+\infty}\id t\,
  \big\langle v(x_t)\cdot \nabla V(x_t) - \nabla\cdot v(x_t)\big\rangle_x^o
\end{equation}
where $\langle \cdot \rangle_x^o$ denotes expectation over the
equilibrium process (that is the process \eqref{mod} with $v=0$)
started from position $x$. (For a mathematical discussion on how
to take the  limits $\varepsilon \to 0$, $T\to +\infty$,
we refer to \cite{mn}.)  It is interesting to recognize here the
linear part in the irreversible entropy flux. When the density of
the test particles is $\mu$ then the instantaneous mean work done
by the background field $v$ is
\[
\caW = \int v\cdot[(v-\nabla V)\mu - \nabla \mu]\,\id x
\]
as the expression between square brackets is the current profile at
density $\mu$.  Clearly, for small driving that equals
\[
\caW = - \int (v\cdot\nabla V - \nabla \cdot v)\,\mu \,\id x + O(\varepsilon^2)
\]
In other words, to linear order in the nonequilibrium background,
$-v(x) \cdot \nabla V(x) + \nabla \cdot v(x)$ equals the mean
dissipated work provided the particle is at $x$.
(Incompressibility of the background fluid can be imposed by
letting $\nabla \cdot v=0$.) That linear term is exactly what
appears in the expectation in
\eqref{ftc3}. Therefore, the linear nonequilibrium correction to the Boltzmann distribution corresponds to the
total dissipated work under the equilibrium relaxation process as started from different initial configurations.

\section{More technical aspects of the nonlocality}\label{ap}

\subsection{Nonlocality in (equilibrium) transient distributions.}

The nonlocal features as discussed in the present paper refer to
fluctuations and responses in steady nonequilibria. Nevertheless,
their origin takes us to Poisson equations for reversible
dynamics, see~\eqref{pois} and \eqref{diff}.  We therefore start
here with a look at equilibrium dynamics but in the transient
regime.

Take a reversible ($v = 0$) diffusion in a potential
landscape $V$ at equilibrium, $\nu = \log\rho = -V$ (up to an
irrelevant constant). Perturb the system by changing the potential
at time $t=0$ to $V + \de V$ and let the system relax towards a
new equilibrium. The evolved distribution at time $t$ be $\mu_t =
\rho + \de\mu_t$, corresponding to the effective time-dependent
potential $\nu + \de\nu_t$, $\de\nu_t = \de\mu_t / \rho$. By the
linear response theory,
\begin{equation}\label{mcl1}
  \frac{\id}{\id t}\, \de\nu_t = L_V \de\nu_t
  + \frac{1}{\rho} \nabla \cdot (\rho \nabla \de V),\quad
  \de\nu_0 = 0
\end{equation}
Using that the second term on the right-hand side equals
\[
  \frac{1}{\rho} \nabla \cdot (\rho \nabla \de V)
  = L_V\,\de V
\]
we find
\begin{equation}
\begin{split}
  \de\nu_t &= \Bigl( \int_0^t e^{s L_V} \id s \Bigr) L_V\,\de V
\\
  &= \bigl( e^{t L_V} - 1 \bigr)\, \de V
\end{split}
\end{equation}
where the equilibrium condition has been used.  As a result,
denoting with $p_t(x,y)$ the transition kernel,
\[
  \de\nu_t(x) = -\de V(x) + \int \id y\, p_t(x,y)\,\de V(y)
\]
or
\begin{equation}\label{nonloctrans}
  \frac{\de\nu_t(x)}{\de V(y)}
  = -\de(x-y) + p_t(x,y)
\end{equation}
Remember that $\de\nu_t$ is only determined up to a constant,
which explains why
$\lim_{t\to\infty} \de\nu_t = -\de V + \langle \de V \rangle_\rho$
differs from the ``naturally expected'' value
$-\de V$; in the above the additive constant has been fixed by the
initial condition $\de\nu_0 = 0$. \emph{The nonlocal part in the linear
response for fixed time $t$ thus exactly equals the transition
probability density $p_t(x,y)$. It is nonlocal in the sense that over distances
where $\rho$ is approximately constant and maximal (or around the minimum of $V$), there is only slow decay in $|x-y|$.  As time grows larger, that effect typically dies out, restoring a  strictly local
response in the infinite-time limit.}

As an example, the
standard one-dimensional Ornstein-Uhlenbeck process (or, oscillator
process) corresponding to the potential
$V(x) = x^2 / 2$
has a response function with the large-time asymptotics
\begin{multline}
  \frac{\de}{\de V(y)}\,\bigl[\nu_t(x) - \nu_t(x')\bigr]
  = \de(x' - y) -\de(x - y)
\\
  + e^{-t} (x'-x) y\, \rho(y) + O(e^{-2 t})
\end{multline}
the nonlocal component of which has a weight exponentially damped in time.

\subsection{Green's function in confinement region}

The response analysis in Sections~\ref{dire}--\ref{invnon} reduces the problem to finding the Green's function
\begin{equation}\label{Green-eq}
  L_\rho^{(x)} \caG(x,y) = -\de(x - y)
\end{equation}
with $L_\rho = \frac{1}{\rho} \nabla \cdot (\rho \nabla)$ generating diffusion in the potential $-\log\rho$. It has an explicit solution in terms of the transition kernel (or transition probability density)):
\begin{equation}\label{Green-explicit}
  \caG(x,y) = \int_0^{+\infty} [p_t(x,y) - \rho(y)]\,\id t
\end{equation}
In a confinement region where $\rho$ is approximately constant around its maximum, one expects that $\caG(x,y)$ is essentially determined by a free diffusion. To asses this conjecture, we need to understand how the inhomogeneities in $\rho$ outside the confinement region influence the Green's function inside it and, more basically, how to make sense to the generally ill-defined expression~\eqref{Green-explicit} for the free diffusion.\\

As well known~\cite{MP}, for a free diffusion (related to the Brownian motion as $\sqrt{2}\,B_t$) in dimension $d$,
with the transition kernel
\begin{equation}
  p_t^\text{free}(0,x) = (4\pi t)^{-d/2} e^{-\frac{|x|^2}{4t}}
\end{equation}
the Green's function $\caG^\text{free}(x) = \int_0^\infty p_t(0,x)\,\id t$
only exists in the transient case, $d \geq 3$. For $d = 1, 2$ the divergence cannot be ``renormalized'' via a stationary density like in~\eqref{Green-explicit} as the latter does not exist. Yet, its divergent part is in fact $x-$independent and one has in \emph{all} dimensions well defined differences
\begin{multline}
  \int_0^\infty [p_t^\text{free}(0,x) - p_t^\text{free}(0,x')]\,\id t
\\
  =
\begin{cases}
  \frac{1}{2} \bigl( |x'| - |x| \bigr) & \text{if } d = 1
\\
  \frac{1}{2\pi} \bigl( \log |x'| - \log |x| \bigr)
  & \text{if } d = 2
\\
  \frac{1}{4} \pi^{-d/2} \Ga(d/2 - 1)
  \bigl( |x|^{2-d} - |x'|^{2-d} \bigr)
  & \text{if } d \geq 3
\end{cases}
\end{multline}
Hence, the free diffusion Green's function is well defined up to a possibly infinite additive constant. However the latter becomes irrelevant due to the ``dipole'' character of the source term in~\eqref{pois} or \eqref{diff}; cf.\ also~\eqref{response-grad} below.

It remains to see in what sense the exterior of a confinement region enters the properties of the (true) Green's function. To simulate that we consider the (standard) diffusion in a cube $[-L/2,L/2]^d$ with reflexive boundary conditions. The transition kernel is
$p_t(x,y) = \prod_{i=1}^d q_t(x_i,y_i)$ with
\begin{multline}
  q_t(x_i,y_i) = \frac{1}{L}
\\
  + \frac{2}{L} \sum_{n \geq 1}
  \cos \Bigl[\pi n \Bigl(\frac{x_i}{L} + \frac{1}{2}\Bigl)\Bigl]
  \cos \Bigl[\pi n \bigl(\frac{y_i}{L} + \frac{1}{2} \bigl)\Bigr]\,
  e^{-\frac{\pi^2 n^2}{L^2}\, t}
\end{multline}
For $d=1$ the Green's function can be obtained explicitly:
\begin{equation}
\begin{split}
  \caG(x,y) &= \int_0^\infty [p_t(x,y) - \rho(y)]\,\id t
\\
  &= \frac{L}{12} - \frac{1}{2}\,|y - x| + \frac{1}{2L}\,(x^2 + y^2)
\end{split}
\end{equation}
Clearly, up to a correction $O(1/L)$ it coincides with the Green's function for free diffusion; moreover, the ``infinite'' additive constant has
been regularized and fixed by the length of the region.

In general one has
\begin{equation}
  p_t(0,x) = \frac{1}{L^d} \sum_k \exp \{
  i k \cdot x - |k|^2 t\}
\end{equation}
with the summation over the dual lattice, $k_i = \ldots, -2\pi /
L, 0, 2\pi / L, \ldots$. Hence,
\begin{equation}
  \caG(0,x) = \frac{1}{L^d}
  \sum_{k \neq 0} \frac{1}{|k|^2}\, e^{i k \cdot x}
\end{equation}
which is to be compared with the free diffusion for which,
formally,
\begin{equation}
  \caG^\text{free}(0,x) = \frac{1}{(2\pi)^d} \int \frac{\id k}{|k|^2}\,
  e^{i k \cdot x}
\end{equation}
The latter is ``infrared divergent'' for $d \leq 2$ and the above
finite lattice version just provides its particular regularization
(above all it provides a cut-off of the neighborhood of $k = 0$).
Note that an alternative (and more standard) way of regularizing the free Green's function is to add a ``positive mass'': for $d = 2$ one obtains
\begin{equation}
\begin{split}
  \caG_\ep(0,x) &= \frac{1}{4 \pi^2} \int \frac{\id k}{|k|^2 + \ep^2}\,
  e^{i k \cdot x}
\\
  &= \frac{1}{2\pi} \int_0^\infty
  \frac{u J_0(u |x|)}{u^2 + \ep^2}\,\id u
\\
  &= \frac{1}{2\pi}\,K_0(\ep |x|)
\end{split}
\end{equation}
with $J_0$ and $K_0$ the Bessel functions of the first
respectively of the second kind. Its short-distance asymptotics is
\begin{equation}
  \caG_\ep(0,x) = \frac{\log 2 - \ga}{2\pi} - \frac{1}{2\pi}\,
  \log (\ep |x|) + o(1)
\end{equation}
($\ga$ being the Euler constant).

These calculations indicate what are the ``boundary effects'' on the Green's function in a region with approximatively constant density profile: the difference from the free Green's function becomes negligible on length scales much smaller than size of the region. Although this comparison includes the removal of an infrared divergence in low dimensions, the latter is well ``renormalizable'' in the above sense of finite differences. Our conclusion is that one could for the present purposes deal just with the free Green's function (although as such being ill defined.)\\

As a specific example we consider the Ornstein-Uhlenbeck process for the diffusion in a quadratic spherically symmetric potential. Its Green's function can be found by solving the equation~\eqref{Green-eq}
with $\rho(x) \propto \exp[-V(x)] \equiv \exp(-\ep^2 |x|^2 / 2)$ (for simplicity we restrict here to a source located at the origin).
Again in two dimension, this has a solution
\begin{equation}
  \caG_\text{OU}(0,x) = -\frac{1}{2\pi} \int_{1/\ep}^{|x|}
  \frac{\id r}{r}\,e^{V(r)}
\end{equation}
up to an arbitrary additive constant. For $\ep |x| \ll 1$, what we
have called the confinement region, it reads
\begin{equation}
  \caG_\text{OU}(0,x) = -\frac{1}{2\pi}\,\log (\ep |x|) + o(1)
\end{equation}
in agreement with either of the two above regularization procedures.
Similarly, for dimensions $d > 2$ one gets
\begin{equation}
  \caG_\text{OU}(0,x) = -\frac{\Ga(d/2)}{2 \pi^{d/2}}
  \int_{1/\ep}^{|x|} \frac{\id r}{r^{d-1}}\,e^{V(r)}
\end{equation}
yielding a power law decay  $~|x|^{2-d}$ if $\ep|x| \ll 1$.

\subsection{``Reciprocity relations'' in the inverse problem}
\label{app: reciprocity}

In contrast to section~\ref{dire}, the analysis of section
\ref{invnon} does not make any use of close-to-equilibrium
assumptions while all the same, in \eqref{diff} and for the
inverse problem, the linear response is given in terms of a
\emph{reversible} process.
A solution to~\eqref{diff} can be written as
\begin{equation}\label{resp}
  \de G(x) = -\int \id y\, \caG(x,y)
  \bigl( \frac{J}{\rho} \cdot \nabla \de\nu \bigr)(y)
\end{equation}
where $\caG(x,y)$ is the Green's function~\eqref{Green-eq}--\eqref{Green-explicit}.
Since the transition probabilities satisfy detailed balance, the Green's function exhibits the same symmetry:
$\rho(x) \caG(x,y) = \rho(y) \caG(y,x)$.

We now propose a partial integration which is allowed if the
current $J$ decays sufficiently fast at infinity; using moreover
that $J$ is divergenceless and that $\de G$ is physically
determined only modulo a constant, one can then write \eqref{resp}
as
\begin{equation}\label{response-grad}
  \de G(x) = \int \id y\, \nabla_y \frac{\caG(x,y)}{\rho(y)} \cdot
  \bigl( J\, \de\nu \bigr)(y)
\end{equation}
and therefore
\begin{equation}\label{reci}
  J(x) \cdot \nabla_x \frac{\de G}{\de\nu(y)}
  = J(y) \cdot \nabla_y \frac{\de G}{\de\nu(x)}
\end{equation}
That relation vaguely resembles an Onsager reciprocity:
$- \frac{\de}{\de\nu(y)} \nabla_x G$
can be interpreted as the extra
(gradient) force at $x$ needed to have a deltafunction-like
response at $y$.

\section{Dynamical fluctuations}\label{dyflu}

Variational principles often arise in the context of fluctuation theory. For example, the entropy and related thermodynamic potentials play a role both as important equilibrium variational functionals (second law) and as functionals governing the equilibrium fluctuations (Einstein's fluctuation theory). Here we sketch how the functionals $\caG$ and $\caI$ introduced in Section~\ref{variational} also fit such a scheme; a more technical account of this problem for overdamped diffusions can be found in~\cite{MNW}.

We consider the diffusion process~\eqref{mod} with the test potential $\Phi$ set to zero. For any random realization $x_t$, $t \geq 0$, of this process we introduce the empirical occupation density
\begin{equation}
  \bar\rho_T(z) = \frac{1}{T} \int_0^T \de(x_t - z)\,\id t
\end{equation}
that counts the relative time spent at each point $z$. Apparently,
$\bar\rho_T(z)$ is the random density dependent on the realized history.
In the limit $T \to \infty$ it converges to the stationary density $\rho$, with probability one by the ergodic theorem. The main result reads that for large but finite times $T$, the probability of fluctuations of $\bar\rho_T$ has the asymptotics given by the large deviation law
\begin{equation}\label{empirical density-LD}
  \bsP( \bar\rho_T = \mu ) \propto e^{-T \caI[\mu]}
\end{equation}
in which the functional $\caI[\mu]$ of Section~\ref{variational} is recognized as an exponential decay rate. Then, the variational inequality
$\caI[\mu] \geq \caI[\rho] = 0$ is a mere consequence of the fluctuation law~\eqref{empirical density-LD}. As observed before, finding $\caI[\mu]$ amounts to solving the inverse stationary problem~\eqref{I-explicit}--\eqref{poisson-again} or, equivalently, to evaluating the variational expression~\eqref{I-DV}.

Similarly, the functional~\eqref{G-explicit} reveals to be the exponential decay rate in the large deviation asymptotics of the joint probability law for the empirical occupation times and the empirical current. For proofs and for more details see~\cite{MNW}.

The large deviation theory for stochastic systems has been started and rigorously established by Donsker and Varadhan, \cite{DV,var}. In the physics literature, these methods go back to the seminal work of Onsager and Machlup, \cite{ons}.


\end{document}